# Experimental Results and Analytical Predictions of EHL Film Thickness

J.-P. Chaomleffel, G. Dalmaz, P. Vergne[*]

*LaMCoS, UMR CNRS/INSA de Lyon 5514*
*Bâtiment Jean d'Alembert, 20 avenue Albert Einstein*
*69621 Villeurbanne cedex, France*

**Abstract**

In this work, we consider several types of lubricants – including non-Newtonian fluids – that were studied under various operating conditions leading us to explore a wide range of dimensionless parameters. The experimental results are compared with predictions given by the usual analytical EHL relationships and by more recently developed models. This broad comparison conducted with particular emphasis on minimum film thickness ($h_m$) showed a fair agreement between experimental data and a few predictions including some obtained from extended models. Commonly used elasto-hydrodynamic lubrication (EHL) models did not systematically gave accurate $h_m$ estimation, whereas minimum film thickness not only is a yield value but also serves as a key parameter in estimating lubrication regimes.

[*] Corresponding author : philippe.vergne@insa-lyon.fr

**NOTATIONS:**

Notations used in the paper are related to circular EHL contacts.

| | |
|---|---|
| $E_d$, $E_b$ | elastic moduli of disk and ball materials |
| $E'$ | reduced elastic modulus of the contacting solids; $2/E' = (1-\nu_d^2)/E_d + (1-\nu_b^2)/E_b$ |
| $G$ | Dowson and Higginson dimensionless material parameter; $G = \alpha.E'$ |
| $H_{DH}$ | Dowson and Higginson dimensionless film thickness; $H_{DH} = h/R$ |
| $H_M$ | Moes dimensionless film thickness; $H_M = h/(R.U^{1/2})$ |
| $h_c$ | central film thickness |
| $h_m$ | minimum film thickness |
| $L$ | Moes dimensionless parameter; $L = G.U^{1/4}$ |
| $M$ | Moes dimensionless parameter; $M = W/U^{3/4}$ |
| $P$ | pressure |
| $R$ | ball radius |
| SRR | slide to roll ratio; $SRR = \Delta u/u_e$ |
| $T$ | temperature |
| $U$ | Dowson and Higginson dimensionless speed parameter; $U = \mu.2.u_e/(E'.R)$ |
| $u_b$ | ball velocity |
| $u_d$ | disk velocity |
| $u_e$ | mean entrainment velocity; $u_e = (u_d+u_b)/2$ |
| $\Delta u$ | sliding velocity; $\Delta u = u_d - u_b$ |
| $W$ | Dowson and Higginson dimensionless load parameter; $W = w/(E'.R^2)$ |
| $w$ | normal applied load |
| $\alpha$ | pressure viscosity coefficient of the lubricant at the inlet temperature |
| $\mu$ | dynamic viscosity of the lubricant at the inlet temperature |
| $\nu_d$, $\nu_b$ | Poisson coefficients of disk and ball materials |
| $\Phi_T$ | film thickness thermal reduction coefficient according to Cheng |
| $\sigma$ | composite RMS roughness of the specimen surfaces |

**SUFFIX:**

| | |
|---|---|
| Ch | Chevalier film thickness prediction |
| exp | experimental results |
| HD | Hamrock-Dowson film thickness prediction |
| M | dimensionless form according to Moes |
| MV | Moes-Venner film thickness prediction |

# INTRODUCTION

Film thickness is probably the first and most important parameter that should be estimated by engineers when designing tribological systems. During the last 10 or 15 years, both experimental and numerical tools used to predict film thickness have been greatly improved, especially in terms of calculation time, spatial resolution, dynamic possibilities, accuracy, etc. However, these advances have probably served to advance understanding of the influence of surface features (roughness, bumps, ridges, dents, etc.) much more than of other issues related to the lubricant itself and its main function; i.e., separation of moving surfaces. Among theses issues is the opportunity of properly accounting for the lubricant behaviour (real lubricants behave mostly as non-Newtonian fluids) or developing engineering tools to predict minimum film thickness ($h_m$) with acceptable confidence over a broad range of operating conditions. This last point is really critical, as $h_m$ not only is used as a yield value but also serves as a key parameter in estimating lubrication regimes by means of the $h_m/\sigma$ ratio.

However, $h_m$ measurements and predictions are absent from most recent papers dealing with film thickness evaluation under severe operating conditions like high loads [1,2] or very thin films [3]. In the domain of conventional elastohydrodynamic lubrication (EHL) film thickness (h > 10 nm), measurements reported by Krupka et al. [4] clearly showed that the slope of the minimum film thickness on a log-log scale was higher than the one predicted by usual EHL models. At the same time, similar deviations have been obtained by Venner [5] in calculations extended to very thin films. A faster decrease with speed was reported for $h_m$ than for the central thickness ($h_c$), and it gave a minimum film thickness gradient close to 0.9 against 0.7 for $h_c$.

In this work, we consider several types of lubricants (mineral & synthetic oils, Newtonian & non-Newtonian fluids) under various operating conditions, leading us to explore a wide range of dimensionless parameters. We chose the fluids not only for their specific properties but also because they are widely used, both as model fluids in laboratory studies and as lubricants in real-life mechanisms. The experimental results are compared with predictions given by the usual analytical EHL relationships and by more recently developed models. In summary, this broad comparison especially dealing with the minimum film thickness enabled us to show a fair agreement between experiments and a few relationships. The extension of an existing model is proposed for the operating conditions at the boundary of the classical EHL domain.

# 1. FILM THICKNESS PREDICTIONS

Many numerical solutions that give pressure distribution and film thickness shape in a smooth EHL contact lubricated with a Newtonian piezo-viscous fluid under fully flooded, isothermal and steady state conditions have been developed during the last 15 years. Among them, the most available analytical solutions for circular contacts are:
- ~The Hamrock–Dowson [6] relationships, which predict central and minimum film thickness as functions of the dimensionless parameters U, G and W;
- ~The Moes–Venner [7] equation, which gives central film thickness as a function of the M and L dimensionless parameters. Compared with U, G and W, the choice of M and L eliminates one dimensionless parameter and allows a graphical representation of the full EHL solutions bounded by the rigid and the elastic isoviscous asymptotic solutions;
- ~The Chevalier [8] central film thickness formula and $h_c/h_m$ ratios as functions of the dimensionless parameters M and L.

In rolling-sliding circular EHL contacts, the minimum film thickness occurs over the side lobes area. As shown by Chevalier (Table 1), the minimum and central film thickness values do not vary in the same way. For increasing load (or the dimensionless parameter M), the minimum film thickness decreases faster than the central film thickness, whatever the considered model or the L value is, as shown in Fig. 1. However, $h_m$ variations obtained from the Hamrock–Dowson equation are lower than those predicted by Chevalier calculated in two different ways. The first one was the direct application of Chevalier's work [8], which gave $h_{m\ Ch}$ in Fig. 1. The second one ($h_{m\ Ch/MV}$ in Fig. 1) uses the $h_c/h_m$ ratios found by Chevalier (Table 1), in which the central film thickness was calculated from the Moes–Venner [7] equation ($h_{c\ MV}$), which is used much more in our field than $h_{c\ Ch}$. Furthermore, it was found [9,10] that the combination of the Chevalier and Moes–Venner models gave the best agreement with measured $h_m$ on different fluids and under various operating conditions. In the following figures, $h_{m\ Ch/MV}$ will be used and named $h_{m\ Ch}$.

Fig. 2 shows the domains considered to establish the above-mentioned models as a function of M and L. Note that the Moes–Venner [7] and Chevalier [8] solutions are based on approximately the same M and L ranges, whereas the Hamrock–Dowson [6] formulae were derived from a much more limited domain. One of the objectives of this study was also to investigate the relevance of these models and to check their domain of validity.

## 2. EXPERIMENTAL DETAILS AND OPERATING CONDITIONS

An EHL ball-on-disc test rig was used for this work. Film thickness measurements were performed on lubricated contacts formed between a transparent disc (made of glass or sapphire) and a steel ball. The specimens, whose properties are reported in Table 2, were driven by two independent brushless motors. Their velocities were controlled with high precision to produce the desired slide-to-roll ratio (SRR).

Balls and discs were carefully polished leading to a composite RMS roughness of the undeformed surfaces lower than 5 nm. The discs were coated on their underside with a thin semireflective chromium layer. The bottom of the ball dips in a test reservoir containing the lubricant, ensuring fully flooded conditions in the contact. The contact, the lubricant and the two shafts that supported the specimens were thermally isolated from the outside and were heated by an external thermal control system. A platinum temperature probe monitored the lubricant temperature in the test reservoir within ± 0.1 °C.

The film thickness measurement technique used in this study is based on differential colorimetric interferometry: it has been detailed and validated to as low as a few nanometres in references [9,10]. The contact area was illuminated with a halogen light source built into the microscope illuminator. The chromatic interferograms produced by the contact were captured by a 3CCD colour video camera and frame-grabbed by a personal computer. The spatial resolution of the captured pictures was close to 1 μm. Compared with the typical contact diameter of several hundreds of micrometres, this provides adequate conditions for an accurate determination of $h_c$ and $h_m$, the latter being determined over very narrow areas at the edges of the contact zone. For measurements under 80 nm, the glass discs were overlaid by a silicon dioxide spacer layer of the same refractive index as the studied lubricant. This technique has been pioneered by Westlake et al. [11] at the end of the 1960s to overcome the major limitations to the classical optical interferometry technique.

The experimental work discussed in this paper covers large ranges of speed, normal load and lubricant properties. Experimental results were plotted as film thickness variations versus the mean entrainment speed $u_e$ and were compared with analytical predictions given by Hamrock and Dowson [6], Nijenbanning, Venner and Moes [7] and Chevalier [8]. The covered M and L ranges, together with lubricant properties, are listed in the figure captions. Most of the fluids investigated here have been extensively studied in this laboratory, and rheological data reported in the figure captions have been actually measured.

The film thickness reduction coefficient ΦT proposed by Cheng [12] is used to only predict the occurrence of significant thermal effects. When ΦT < 1, the lubricant properties are modified by shear heating, whereas if ΦT is close to 1, the lubricant flow remains isothermal. From in situ measurements of both pressure and film thickness, Jubault et al. [13] showed that shear heating appeared in the contact inlet when ΦT ≤ 0.96. This limit is plotted by a vertical full line in the figures. A second vertical but dotted line is plotted when the maximum L value in the Chevalier model is exceeded (L > 20).

## 3. RESULTS AND DISCUSSION

The study is organized according to the lubricant's rheological behaviour and the contact pressure range experienced during the experiments.

Firstly, mineral paraffinic base oil, Squalane and Pennzane, studied under moderate contact pressure, are considered as Newtonian fluids. Squalane is a known reference fluid in EHL thin film studies and in non-equilibrium molecular dynamic simulations of confined films. Pennzane is a synthetic hydrocarbon that nowadays competes with perfluoropolyether fluids in spatial lubricated mechanisms. The mineral paraffinic lubricant is representative of base oils used in numerous industrial applications.

Then, the non-Newtonian behaviour of Z25 and PAO650 under moderate pressure is discussed. The former is well known in the space industry and the latter is similar to polymeric additives used in automotive applications, for instance.

We chose 5P4E to investigate the contact pressure influence on film thickness. This synthetic fluid has been extensively studied in the community from the 1960s to the 1980s, and a large number of papers involving this fluid have already been published.

Finally, we used Santotrac 50 to combine non-Newtonian behaviour and high contact pressure. Santotrac 50 is a traction fluid, and there is renewed interest in these lubricants, which are used in continuously variable transmissions.

### 3.1 Newtonian behaviour at moderate contact pressure

Three hydrocarbons are firstly considered: a mineral paraffinic base oil, Squalane and a synthetic hydrocarbon of the PAO12 type named Pennzane. From literature results and considering their structure, it is believed that these lubricants behave as Newtonian fluids under ambient conditions. However, we reported [10] the occurrence of immobile boundary layers at the specimen surfaces for Squalane. Accordingly, the layer thickness (≈1.7 nm) has been subtracted from the measured values to report only the viscous hydrodynamic contribution to film thickness formation in Fig. 5.

Concerning $h_c$, it is clearly visible from Figs. 3–5 that both the Hamrock–Dowson [6] and Moes–Venner [7] relationships give acceptable central film thickness predictions down to a few nanometres. As a confirmation, linear regressions carried out to determine the $h_c$ slope in log-log plots over the entire sets of data measured under isothermal conditions gave 0.68, 0.66 and 0.66 for mineral base oil, Squalane and Pennzane respectively, as the speed exponent in the Hamrock–Dowson relationship equals 0.67. This also confirms why it is well accepted to extend $h_{c\,HD}$ over a much broader range than the initial domain used by Hamrock and Dowson to develop their formula.

However, similar extrapolations applied to $h_{m\,HD}$ can lead to large deviations. From Figs. 3–5, it is easy to note that the Hamrock–Dowson [6] formula for $h_m$ is unable to predict minimum film thickness within an acceptable accuracy over a large range of operating conditions. The gradients calculated from linear regressions performed on measured minimum film thickness values in the isothermal domain, i.e., 0.75, 0.81 and 0.85 for the mineral base oil, Squalane and Pennzane respectively, do not coincide with the value of 0.68 proposed in [6]. Moreover, the Hamrock–Dowson model systematically overestimates $h_m$, which could lead to hazardous situations, especially for tribological systems working under a very-thin-film EHL regime, corresponding to high M and low L conditions. This regime is noted as VTF (very thin films) in the chart of Fig. 2.

A better agreement was found using Chevalier's $h_c/h_m$ ratios as explained before. The corresponding $h_{m\,Ch}$ gradients vary from 0.82 to 0.85 and thus are in a much closer range to the result obtained experimentally than that using the Hamrock–Dowson relationship. However, Chevalier's ratios are limited to $M \leq 1000$ (Table 1), and attempts to extrapolate values outside this boundary appeared very questionable for several reasons.

Compared with $h_c$, $h_m$ occurs over a very small area. Whether the technique is numerical or experimental, its evaluation could be tricky and could depend on mesh size or on spatial resolution. Furthermore, $h_m$ variations versus M and L are difficult to model. The physical mechanisms that govern $h_m$ are less understood than those involved in $h_c$.

Another reason concerns the ultimate $h_c/h_m$ ratios; i.e., when L is low and M becomes higher than 1000. Experimental results (see Figs. 3–5) and numerical models (Chevalier [8], but also Venner [5]) obtained so far suggest increasing $h_c/h_m$ ratios when M increases at constant L. However, it is impossible to affirm whether the actual $h_c/h_m$ values would continue to increase or not.

The continuum mechanics hypothesis is no longer valid when numerical simulations are performed at the nanometre or subnanometre scales. Film thickness is of the same dimension as molecular size, and the significance of results becomes questionable. Moreover, experimental data are limited by both spatial and thickness resolutions.

Relative deviations between calculated and experimental film thickness values for Pennzane are reported in Fig. 6. Under isothermal conditions ($u_e < 1$ m/s) $h_c$ and $h_m$ obtained from the Hamrock–Dowson formula and from Chevalier's ratios respectively agree within 5–10% with measured film thickness values, whereas $h_m$ deduced from Hamrock and Dowson [6] overstates the actual values. Furthermore, the occurrence of relatively high thermal effects produces a change in the slope of the measured film thickness (Fig. 5) and thus an increase of the relative deviations (Fig. 6), which vary in the same way when the isothermal boundary is exceeded.

### 3.2 Non-Newtonian behaviour at moderate contact pressure

Two different non-Newtonian synthetic lubricants are concerned, a linear perfluoropolyalkylether (Z25, ca. 15000 kg/kmole) and a high-molecular-weight polyalphaolefin (PAO650, ca. 20000 kg/kmole). The shear-thinning behaviour of both fluids was proved by rheological tests performed with high-pressure viscometers and confirmed by in-contact measurements conducted to evaluate its influence on film thickness [14–16].

The Z25 results (Fig. 7) reveal a moderate difference from predictions based on Newtonian behaviour. Actually, experimental $h_c$ and $h_m$ slopes are lower, respectively 0.59 and 0.74, and an increasing deviation between measured and predicted film thickness occurs especially at moderate and high speeds. On the other hand, one can notice that measured and calculated film thickness are very close when $ue \leq 0.01$ m/s.

It is deduced from these observations that the Z25 response is typical of a moderate shear thinning behaviour. High shear stress measurements [14] showed that the Z25 generalized viscosity decreases with shear stress following a Carreau model with an exponent of 0.82. This behaviour could be expected from an approximate Newtonian shear stress limit estimation based on the fluid molecular weight. Further experiments [16] conducted at variable slide-to-roll ratios indicated a continuous decrease of both $h_c$ and $h_m$ with increasing SRR. They firstly confirmed that the experimental data plotted in Fig. 7 corresponded to the first non-Newtonian transition, and secondly, they showed that even under pure sliding conditions, a second Newtonian plateau was not achieved.

More interestingly, the rheological effects reported above are enhanced when considering PAO650 (Fig. 8). Compared with Z25, this fluid possesses a lower critical stress for shear thinning, a lower Carreau exponent (0.74), a higher viscosity at ambient temperature and a pressure viscosity coefficient close to that of commonly used lubricants.

Even if $h_c$ and $h_m$ slopes remain quite close to those derived from Fig. 7 (0.59–0.70 against 0.59–0.74), an important shift is found between the measured and predicted film thickness values. It must be noticed that the operating conditions experienced with PAO650 are situated inside the M and L domains considered when classical EHL models were established. Thus, the large deviations observed in Fig. 8 can only be attributed to the non-Newtonian behaviour of PAO650. It was shown [15] that central and minimum film thickness values are well described using the Carreau relationship. Film thickness is thus entirely predictable from the rheological properties obtained from viscometers using simple calculations. This proves that shear-thinning, occurring mainly in the contact inlet, is the dominant effect on the shearing response of this fluid, in the absence of measurable heating.

In case of marked non-Newtonian effects as with PAO650, it is of major importance to have a realistic estimation on how the film thickness distribution is affected. Fig. 9 answers this issue: relative deviations between calculated and experimental film thickness values are plotted. $h_{m\ HD}$ estimation (only reported for comparison) gives an almost constant relative deviation when the mean entrainment speed varies. From $h_{c\ HD}$ and $h_{m\ Ch}$ calculations that have been proved to predict film thickness accurately for Newtonian lubricants, it is clearly shown that increasing $u_e$ leads to an increasing gap between Newtonian predictions and measurements. However, it is also noteworthy that the considered relative film thickness deviations exhibit virtually identical slopes, proving that $h_c$ and $h_m$ are similarly influenced by shear thinning. This confirms that the lubricant properties in the inlet region remain the key parameters for the film thickness building up, as in the Newtonian case.

### 3.3 Newtonian behaviour at high contact pressure

So far, we have considered low EHL contact pressures because experiments were carried out with glass discs. In the following stage, contact pressure has been extended to higher values using sapphire discs that permit much higher normal loads and thus higher pressures to be sustained.

5P4E polyphenyl ether (or m-bis(m-phenoxyphenoxy)benzene) has been studied [13] over large ranges of normal loads and speed conditions. This fluid has been extensively studied and has been considered as a model lubricant by the community since the 1970s. The combination of its physical properties with the operating conditions applied here leads to M and L dimensionless parameters somewhat beyond the classical EHL ranges: for instance at 50 °C, L is larger than 20 when $u_e \geq 0.35$ m/s, and M is smaller than 10 when $u_e$ is large and w (the normal load) is low as well. These "extreme" conditions lead to highly loaded thick films that are totally different from the conditions discussed in the first part of section 1. They are noted as HLTF (highly loaded thick films) in Fig. 2, an area that covers the thick EHL regime together with the transition toward the isoviscous rigid asymptote, therefore toward the hydrodynamic regime.

Fig. 10 reports both predicted and measured $h_c$ and $h_m$ for three loads and seven entrainment speeds. In most cases, $h_c$ is estimated within 10% and even within 5% for the majority of operating conditions, whatever the considered analytical model, Hamrock–Dowson [6] or Moes–Venner [7]. The worse case has been encountered applying the later model with the maximum normal load: a significant overestimation of $h_c$ has been found and can also explain the more visible (Fig. 10) $h_{m\ CH}$ overestimation at P = 1.8 GPa obtained using Chevalier's $h_c/h_m$ ratios. Actually, this deviation could be already predictable from Fig. 1b) where for L = 20, $h_{c\ MV}$ is larger than $h_{c\ HD}$ or $h_{c\ CH}$, whatever the M value.

In spite of this discrepancy, we preferred to apply the combination of Chevalier and Moes–Venner models—they gave the best agreement under low contact pressures—for the sake of consistency and homogeneity of the results discussed in this work. For this purpose, we applied an interpolation procedure based on linear combinations of Lagrange polynomials on the original $h_c/h_m$ ratios from [8]. Then, we extended the Chevalier M and L domain by extrapolating using the polynomial functions to L = 40 and M = 3. The extended Chevalier table is reported in Table 3 and has been used for the $h_{m\ CH}$ calculations plotted in Fig. 10.

Experimental results show increasing $h_m$ versus $u_e$ slopes when the normal load, and thus the contact pressure, increase. Experimentally, we find gradients of 0.74, 0.87 and 0.99 for 0.6, 1.1 and 1.8 GPa respectively, whereas the Hamrock–Dowson model gives a constant slope of 0.68.

Because the estimation of central film thickness was not totally satisfying for a given application, we used Chevalier's work to compare our experimental data with predicted values, especially on minimum film thickness. Fig. 11 presents $h_c/h_m$ ratios obtained from the Hamrock–Dowson [6] and extended Chevalier models (Table 3) and from tests. The first model gives an almost constant ratio that refutes most of the experimental findings. We explained before that this was due to the inability of the Hamrock–Dowson minimum film thickness relationship to produce accurate values. On the other hand, the agreement between predictions from the extended Chevalier table and experimental results is indeed very consistent and justifies our initial choices about this model. $h_c/h_m$ varies over a range from 1.25 to more than 3, according to the contact load and the mean entrainment speed values. This confirms that knowledge of $h_c$ only does not represent a realistic view of the film thickness distribution: areas of much lower film thickness can exist and can present potential risk of direct contact between solid body surfaces and thus damage. Furthermore, the similarity between the two sets of data remains valid for the two highest speed conditions; i.e., conditions that generate shear heating. The latter is likely to occur in the contact inlet and is supposed to affect $h_c$ and $h_m$ in the same way, as the experimental ratios agree with the theoretical and thus isothermal ones.

**3.4 Non-Newtonian behaviour at high contact pressure**

To complete this study, it was necessary to discuss the case of highly loaded contacts lubricated by non-Newtonian fluids. For this purpose, we chose the most widely used traction fluid, Santotrac 50 composed of a synthetic base oil (Santotrac 40, dicyclohexyl alkane type, very-high-pressure viscosity coefficient) and a high-molecular-weight polymeric additive that induces a shear thinning behaviour of the mixture at low shear stress. It was shown that Santotrac 40 behaves as a Newtonian fluid whose viscosity at 25 °C is approximately two-thirds that of the Santotrac 50 viscosity at the same temperature [17]. It was also proved that both lubricants have equal pressure viscosity coefficients, because of the small amount of polymer in Santotrac 50 (a few % w/w).

In Figs. 12 and 13, experimental results obtained at 25 °C and 1.3 GPa are presented, together with predictions based on Santotrac 50 and Santotrac 40 data respectively. We used the extended Chevalier table (Table 3) to calculate $h_{m\ Ch}$ because L exceeds 20 when $u_e \geq 0.4 m/s$. These experiments covered the medium M to high L domain. For the purpose of comparison with previously published data, the film thickness is plotted in the Dowson and Higginson dimensionless form.

As expected, the measured values deviate significantly from simulated Santotrac 50 film thickness values (Fig. 12). Nevertheless, the central film thickness gradient remains almost equal to the Hamrock–Dowson U exponent, although the experimental $h_m$ gradient is close to 0.82. These gradients are very similar to those reported for Newtonian fluids under moderate contact pressure (section 3.1). Here, because of the shear thinning behaviour of the polymeric additive the non-Newtonian response of Santotrac 50 leads to a whole reduction of the actual film thickness values compared to the predictions. This global shift is confirmed in Fig. 13, where the theoretical results are obtained from the Santotrac 40 rheological properties; i.e., totally ignoring the polymer influence. Under this condition, the agreement between measurements and analytical results appears much more satisfying, and the experimental points are now surrounded by predicted film thickness values (Fig. 13). All this occurs as if a second Newtonian plateau occurs in the actual lubricant behaviour, which would be very close to that of the base oil. A deviation of around 10% persists, but it was not our intention to speculate on the accurate value of the actual lubricant viscosity during the tests. Lubricant composition being subject to variations from batch to batch, the ratio of 2/3 taken from [17] has been considered just as a typical value.

## **CONCLUSIONS**

This paper has aimed to point out how film thickness could be influenced by lubricant behaviour and operating conditions, and how far analytical models were able to predict $h_c$ and $h_m$.

Our experimental results show that central film thickness could be accurately calculated from the Hamrock–Dowson model [6] well outside the range of the conditions considered when it was proposed. This point has also been confirmed by recent numerical papers [18]. Moreover, the Moes–Venner [7] relationship gives an acceptable estimation of $h_c$, especially when the L dimensionless parameter is low.

On the other hand, the Hamrock–Dowson formula is not able to fit properly the minimum film thickness data, whatever the operating conditions. A much better agreement has been found applying the work of Chevalier [8], who published tabulated $h_c/h_m$ ratios as a function of the dimensionless parameters M and L. The quantitative comparison between experimental results and analytical predictions has been successfully conducted outside the models domain of validity thanks to the extension of Chevalier's table to M = 3 and L = 40.

The occurrence of shear thinning reduces the slopes of the film thickness versus mean entrainment velocity results when the effective lubricant viscosity does not reach a stable value in the contact inlet (Z25 and PAO650). However, a global shift toward lower film thickness while keeping the slope constant was observed when a second Newtonian plateau was achieved.

Nowadays, elastohydrodynamic films are of nanometre rather than micrometre proportions. This has been possible thanks to numerous contributions on film thickness build-up mechanisms published during the last 30 years. Under very thin film thickness—i.e., when simultaneously M becomes higher than 1000 and L remains lower than 2—both experimental results and numerical models [5,8] suggest an increasing $h_c/h_m$ ratio. However, so far it has been impossible to predict whether in real contact the ratio continues to increase or whether it follows another tendency. Ignoring the continuum mechanics hypothesis and thanks to progress in computer modelling, numerical simulations can be conducted down to the Ångstrom scale. However, keeping in mind the physical reality of thin-film EHL contacts, a different behaviour was expected. At least a criterion taking into account the operating conditions and the lubricant molecular dimensions could be proposed as a boundary. However, a better understanding of the mechanisms that influence $h_m$ is probably necessary to achieve this advance. It would be useful to pursue this endeavour to develop a minimum film thickness model valid in the high M (above 1000) low L (below 2) thin film domain.

| $h_c/h_m$ | | M | | | | |
|---|---|---|---|---|---|---|
| | | 10 | 30 | 100 | 300 | 1000 |
| L | 0 | 1.26 | 1.25 | 1.33 | 1.48 | 1.93 |
| | 2 | 1.35 | 1.48 | 1.80 | 2.23 | 3.28 |
| | 5 | 1.35 | 1.57 | 1.92 | 2.42 | 3.43 |
| | 10 | 1.35 | 1.54 | 1.87 | 2.33 | 3.20 |
| | 20 | 1.31 | 1.46 | 1.72 | 2.08 | 2.79 |

Table 1: Central versus minimum film thickness ratios according to Chevalier [8]

| | Glass disk | Steel ball | Sapphire disk |
|---|---|---|---|
| Elastic modulus | 81 GPa | 210 GPa | 360 GPa |
| Poisson ratio | 0.208 | 0.3 | 0.34 |
| Reduced elastic modulus | E' = 124 GPa | | E' = 295 GPa |

Table 2: Elastic properties of the contacting materials

| $h_c/h_m$ | | M | | | | | |
|---|---|---|---|---|---|---|---|
| | | 3 | 10 | 30 | 100 | 300 | 1000 |
| L | 0 | 1.26 | 1.26 | 1.25 | 1.33 | 1.48 | 1.93 |
| | 2 | 1.31 | 1.35 | 1.48 | 1.80 | 2.23 | 3.28 |
| | 5 | 1.27 | 1.35 | 1.57 | 1.92 | 2.42 | 3.43 |
| | 10 | 1.28 | 1.35 | 1.54 | 1.87 | 2.33 | 3.20 |
| | 20 | 1.26 | 1.31 | 1.46 | 1.72 | 2.08 | 2.79 |
| | 40 | 1.21 | 1.23 | 1.30 | 1.42 | 1.58 | 1.97 |

Table 3: Extended Chevalier table to L = 40 and M = 3

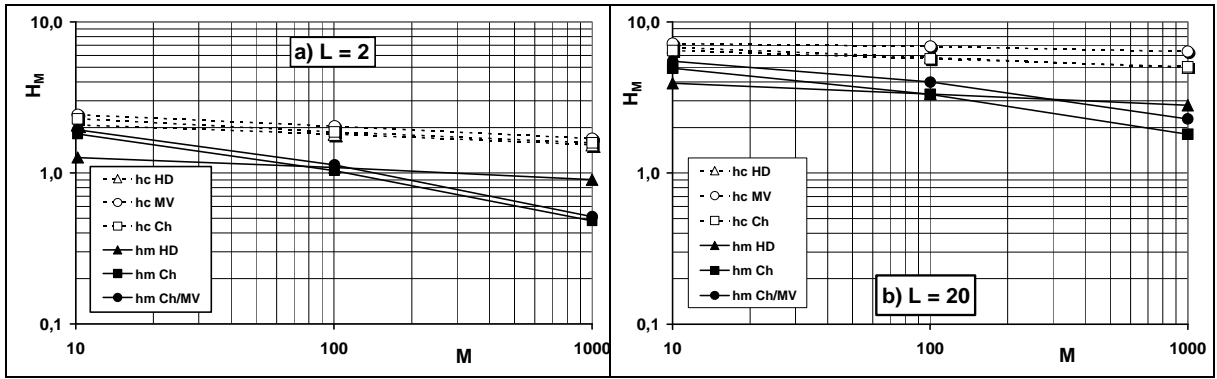

Figure 1: Predicted dimensionless film thicknesses for a) L = 2 and b) L = 20

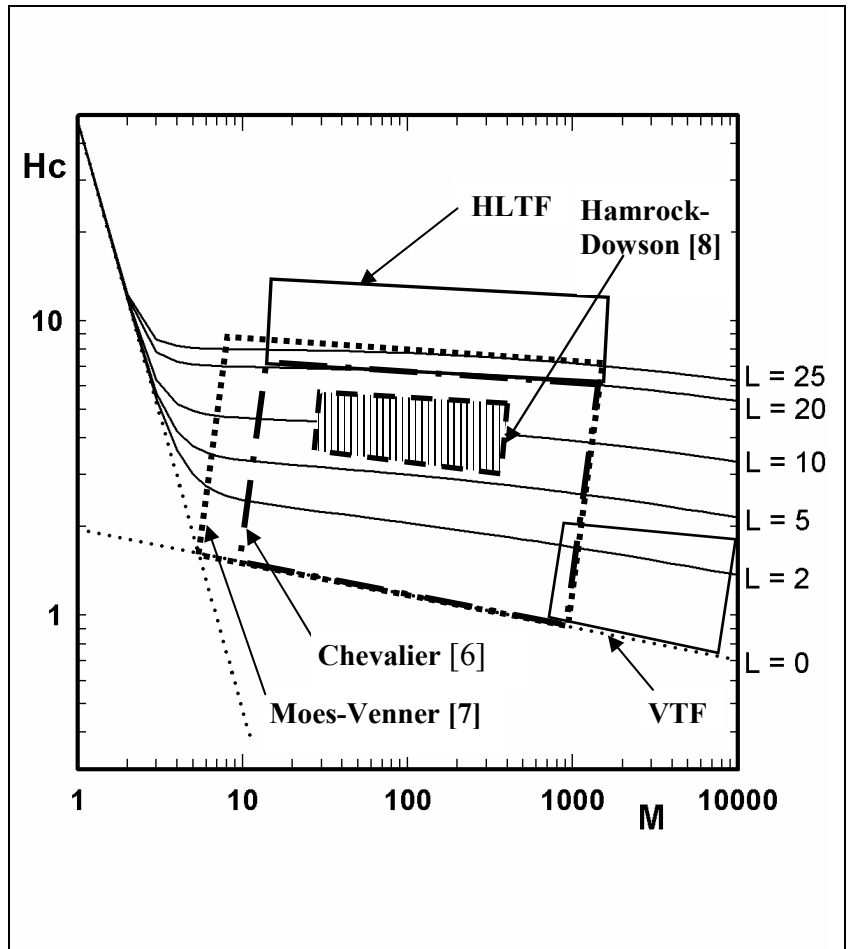

Figure 2: Validity domains of film thickness models
VTF = Very Thin Films ; HLTF = Highly Loaded Thick Films

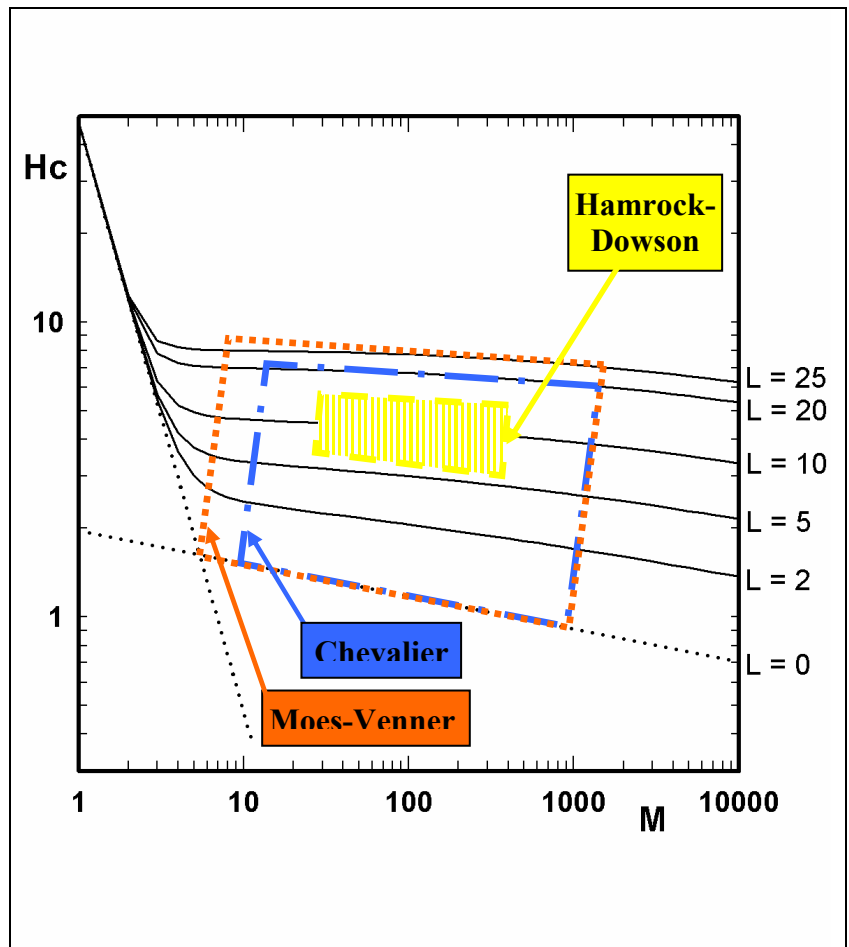

Figure 2: Validity domains of film thickness models
VTF = Very Thin Films ; HLTF = Highly Loaded Thick Films

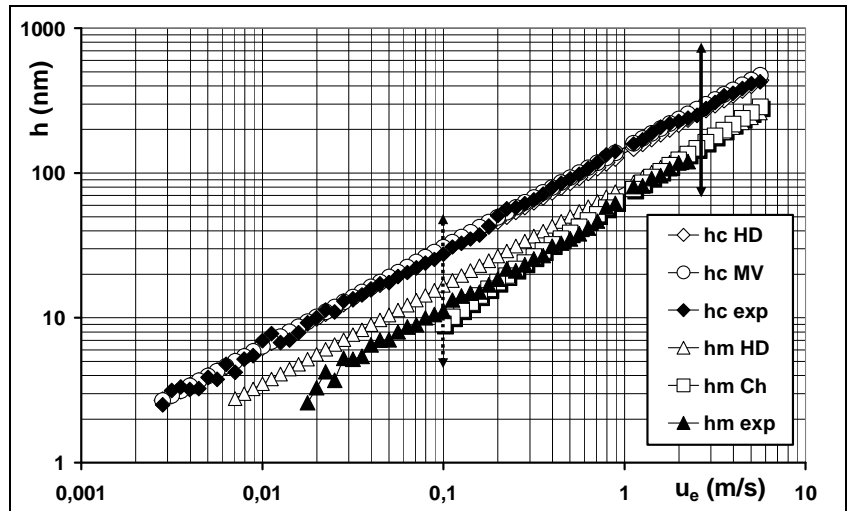

Figure 3: Film thickness versus $u_e$ for a mineral oil
(M = 15000; L = 1.25 to M = 50.7; L = 8.4)
*T = 50 °C, $P_H$ = 0.5 GPa, $\mu$ = 10.3 mPa.s, $\alpha$ = 22.5 GPa$^{-1}$*

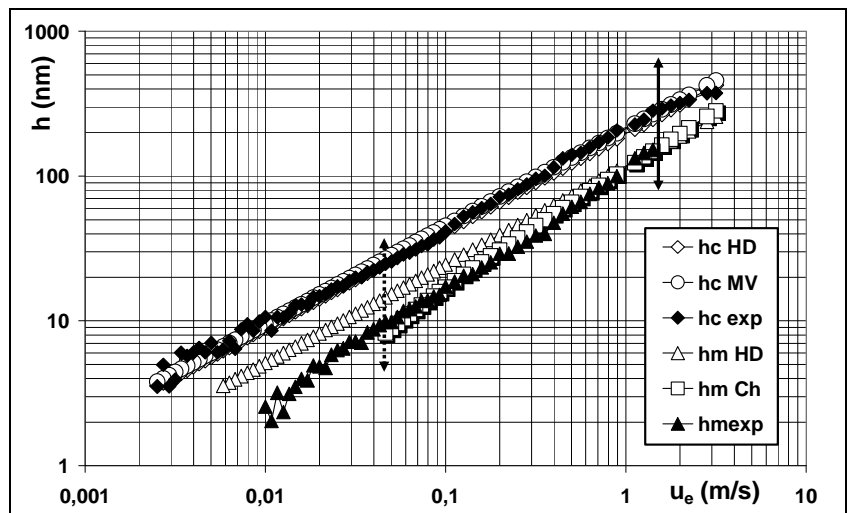

Figure 4: Film thickness versus $u_e$ for Squalane
(M = 8800; L = 1.0 to M = 41.8; L = 6.1)
*T = 25 °C, $P_H$ = 0.5 GPa, $\mu$ = 26 mPa.s, $\alpha$ = 15.4 GPa$^{-1}$*

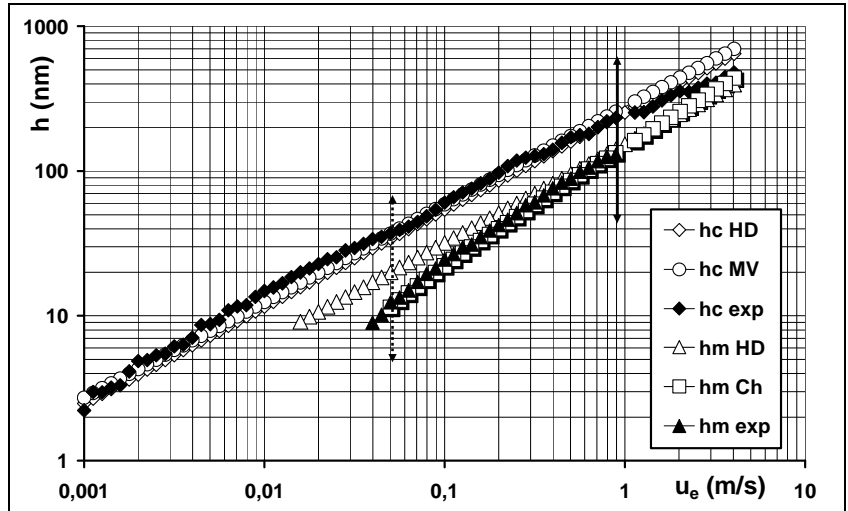

Figure 5: Film thickness versus $u_e$ for Pennzane
(M = 18200; L = 0.8 to M = 35.9; L = 6.2)
*T = 60 °C, $P_H$ = 0.51 GPa, μ = 40.3 mPa.s, α = 14.6 $GPa^{-1}$*

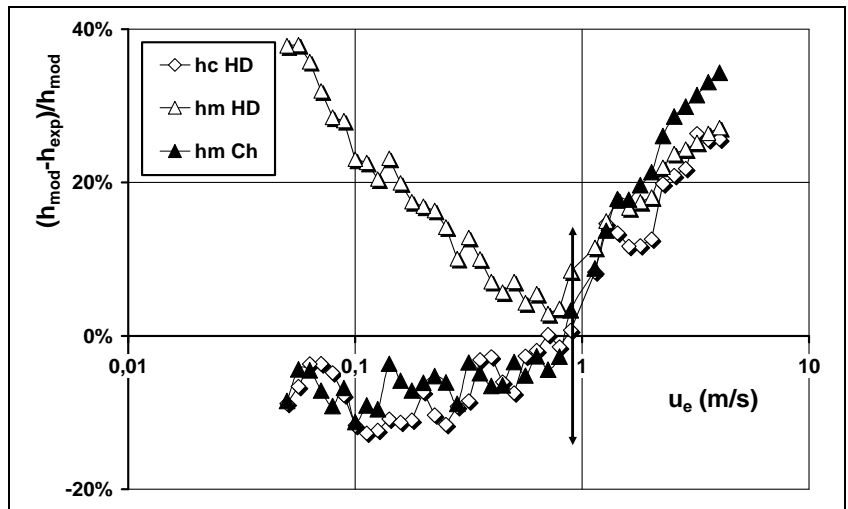

Figure 6: Relative film thickness deviations for Pennzane

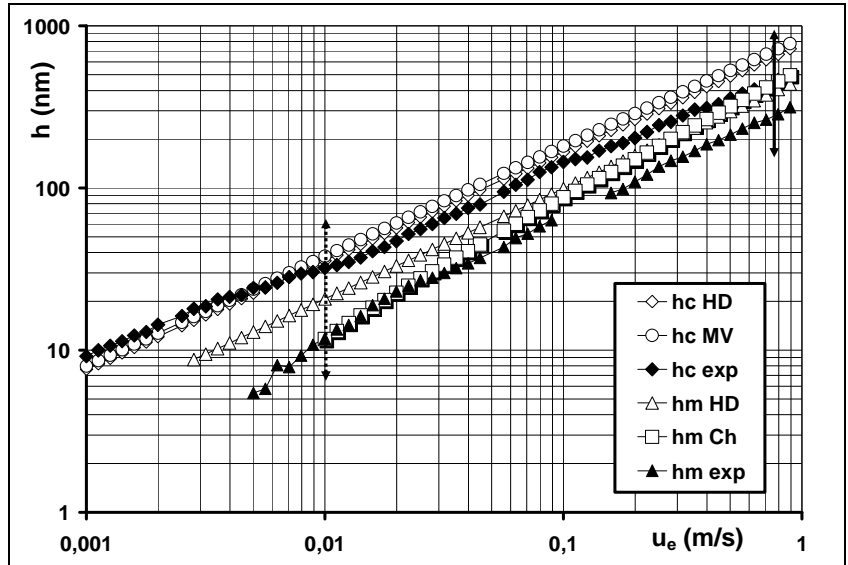

Figure 7: Film thickness versus $u_e$ for Z25
(M = 5400; L = 1.4 to M = 33.3; L = 7.6)
*T = 60 °C, P = 0.5 GPa, μ = 0.18 Pa.s, α = 17.5 GPa$^{-1}$*

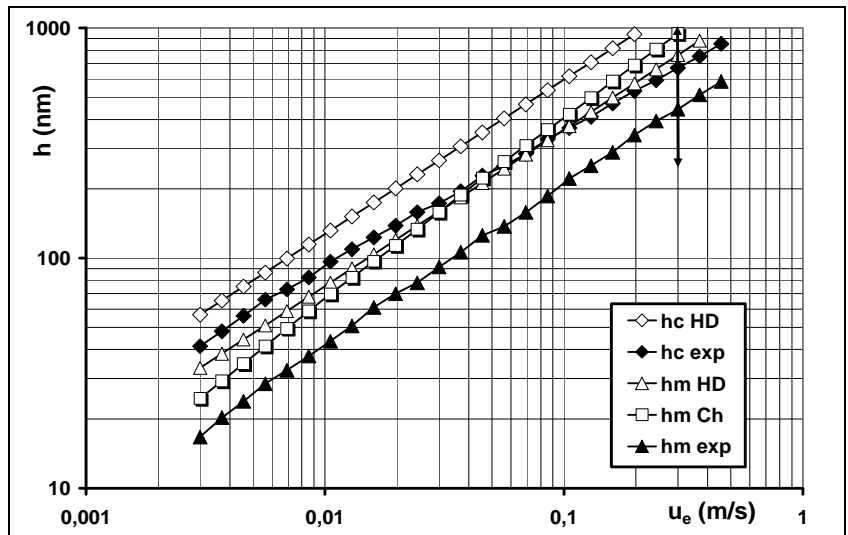

Figure 8: Film thickness versus $u_e$ for PAO650
(M = 451; L = 2.8 to M = 10.4; L = 9.8)
*T = 75 °C, $P_H$ = 0.53 GPa, μ = 1.42 Pa.s, α = 14.8 GPa$^{-1}$*

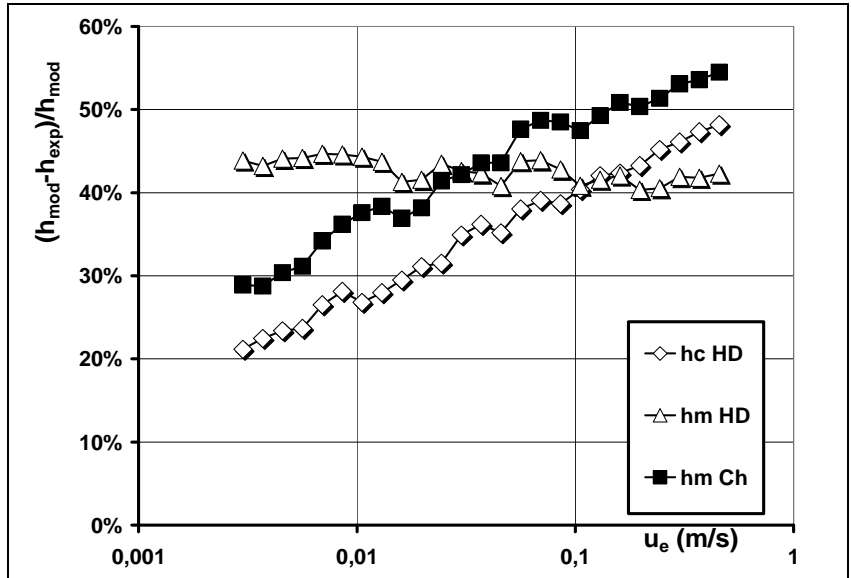

Figure 9: Relative film thickness deviations for PAO650

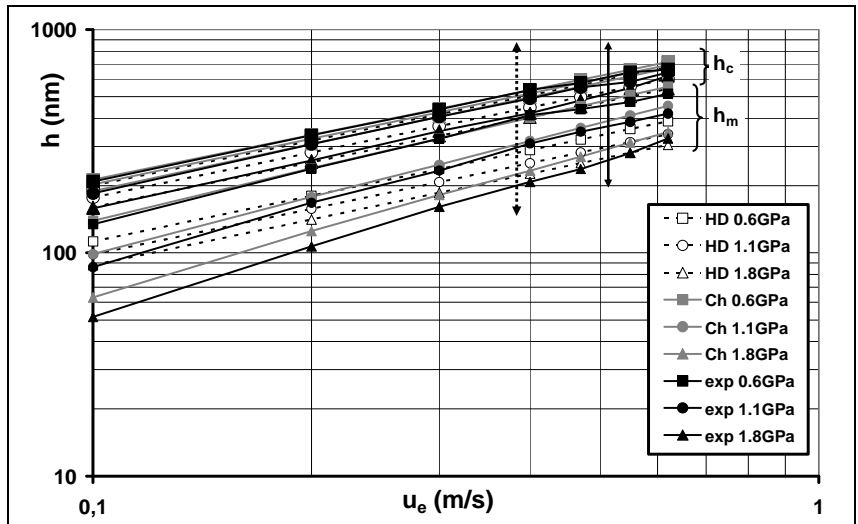

Figure 10: Film thickness versus $u_e$ for 5P4E
(M = 980; L = 14.3 to M = 8.9; L = 22.6)
*T = 50 °C, μ = 0.16 Pa.s, α = 28.4 GPa$^{-1}$*

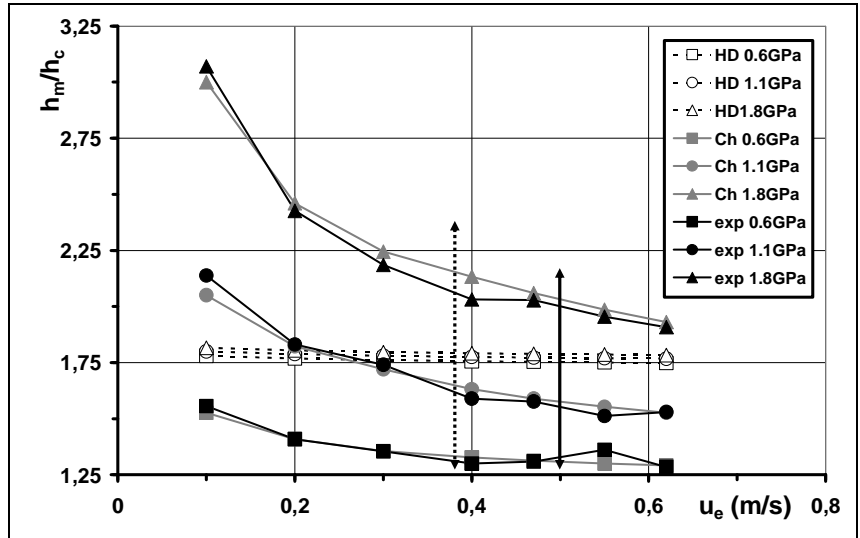

Figure 11: $h_c/h_m$ ratios for 5P4E at T = 50 °C

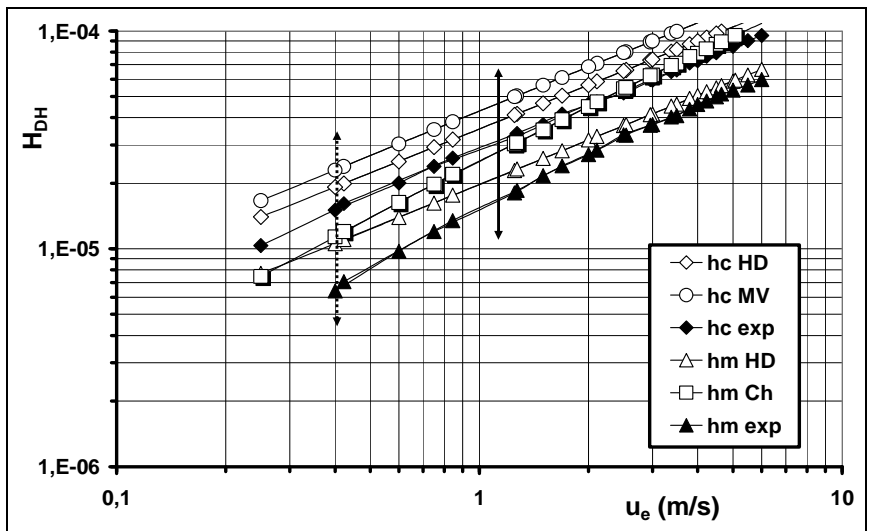

Figure 12: Dimensionless film thickness versus $u_e$ for Santotrac 50
(M = 378.2; L = 17.5 to M = 34.9; L = 38.8)
*T = 25 °C, P = 1.3 GPa, µ = 56 mPa.s, α = 36 GPa$^{-1}$*

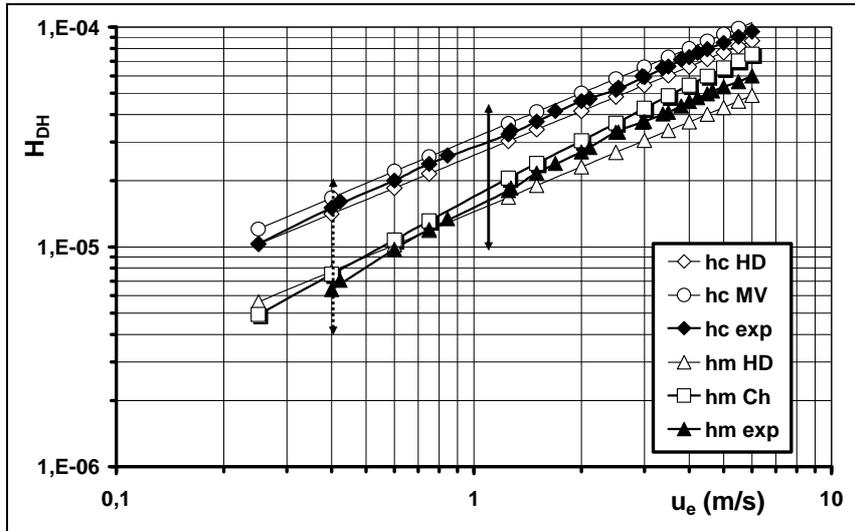

Figure 13: Comparison between dimensionless simulated Santotrac 40 film thicknesses and measured values on Santotrac 50
*T = 25 °C, P = 1.3 GPa; Santotrac 40: $\mu$ = 37 mPa.s; $\alpha$ = 36 GPa$^{-1}$*